\documentclass[cits]{PoS}

\usepackage{amsmath,amssymb}
\usepackage{dsfont}
\usepackage{bm}
\usepackage{tikz}
\usetikzlibrary{calc,matrix,arrows,decorations.pathmorphing,decorations.markings}

\newcommand{\bvec}[1]{{\ensuremath{\mathbf{\bm{#1}} }}}
\newcommand{\Dif}{\text{d}}
\newcommand{\braket}[3]{\ensuremath{\left< #1 \vphantom{#3} \right| #2 \left| #3 \vphantom{#1} \right>}}

\title{Evolution and dynamics of cusped light-like Wilson loops}

\ShortTitle{Evolution and dynamics of cusped light-like Wilson loops}

\author{\speaker{Frederik Van der Veken}%
\\
       University of Antwerp\\
       E-mail: \email{frederik.vanderveken@ua.ac.be}}

\abstract{
We address a connection between the energy evolution of the polygonal light-like Wilson exponentials and the geometry of the loop space with the gauge invariant Wilson loops of a variety of shapes being the fundamental degrees of freedom. The renormalization properties and the differential area evolution of these Wilson polygons are studied by making use of the universal Schwinger quantum dynamical approach. We discuss the appropriateness of the dynamical differential equations in the loop space to the study of the energy evolution of the collinear and transverse-momentum dependent parton distribution functions.
}

\FullConference{XV International Conference on Hadron Spectroscopy-Hadron 2013\\
		4-8 November 2013\\
		Nara, Japan }

\begin{document}

\section{Introduction}
A standard transverse momentum dependent parton density function (or TMD for short) can be defined as:
\begin{equation}
	f(x,\bvec{k}_\perp) = 
		\frac{1}{2}\int \frac{\Dif z^-\Dif^2 \bvec{z}_\perp}{2\pi(2\pi)^2} \; e^{ik\cdot z}
		\braket{P,S}{\bar{\psi}(z) \, U^\dag (z;\infty) \Gamma \, U(\infty;0) \, \psi(0)}{P,S} \Big|_{z^+=0}
\end{equation}
where the Wilson lines are split into their longitudinal and transversal parts:
\begin{align}
	U(\infty, 0) &=
		U(\infty^-,\bvec{\infty}_\perp ; \infty^-, \bvec{0}_\perp) U(\infty^-,\bvec{0}_\perp ; 0^-, \bvec{0}_\perp)\\
	&=
		\mathcal{P} \,\text{exp}\left[-i g \int_{0}^{\infty} \Dif z_\perp \; A_\perp(\infty^-, \bvec{z}_\perp)\right]
		\mathcal{P} \,\text{exp}\left[-i g \int_{0}^{\infty} \Dif z^- \; A^+(z^-, \bvec{0}_\perp)\right].
		\nonumber
\end{align}
For a general discussion on TMDs and Wilson lines see \cite{TMDdef1,TMDdef2,TMDdef3,TMDdef4}. TMDs have a complex singularity structure, and especially light-like singularities, coming from segments that lie on the light-cone, are difficult to treat (see \cite{llTMD}). Overlapping divergences will manifest themselves as terms of the order $1/\epsilon^2$, where $\epsilon$ is the regulator used in dimensional regularisation. These will give the only contribution to the evolution equations, which are governed by the cusp anomalous dimension. The latter is given by \cite{cad1,cad2,cad3}
\begin{equation}
\label{eq:cad}
	\Gamma_{\text{cusp}} = \frac{\alpha_s \, C_F}{\pi} \left( \chi \coth \chi - 1 \right)
	\quad \stackrel{\text{on-LC}}{\longrightarrow} \quad
	\frac{\alpha_s \, C_F}{\pi}
\end{equation}
where $\chi$ is the cusp angle. In the on-LC limit we have $\chi\rightarrow\infty$, making the cusp anomalous dimension independent on $\chi$ (often referred to as a `hidden cusp').\footnote{Actually it becomes infinite, but will be renormalised to the factor in front of the double pole.}

\section{Wilson loops and loop space}
A Wilson loop is a trace of a Wilson line on a closed path, evaluated in the ground state:
\begin{equation}\label{eq:wloop}
	\mathcal{W}[C] =
		\frac{1}{N_c}\,\text{tr}\, \braket{0}{\mathcal{P} \,\text{exp}\left[i g \oint_C \Dif z^\mu A^a_\mu (z) t_a \right]}{0}
\end{equation}
where $C$ is any closed path and $A_\mu^a$ is the (non-Abelian) gauge field. This loop is a pure phase, transforming coordinate dependence into path dependence. By extending the definition of Wilson loops, they can be used as elementary building bricks to completely recast QCD in loop space \cite{QCDrecast1,QCDrecast2,QCDrecast3}. A $n$-th order Wilson loop (consisting of $n$ sub-loops) then becomes
\begin{align}
	\mathcal{W}_n(C_1,\ldots, C_n) &=
		\braket{0}{\Phi(C_1) \ldots \Phi(C_n)}{0}, \\
	\Phi (C) &=
		\frac{1}{N_c}\,\text{tr}\, \, \mathcal{P} \,\text{exp}\left[i g \oint_{C} \Dif z^\mu A_\mu (z) \right].
\end{align}
Note that all gauge kinematics are encoded in a $\mathcal{W}_1$ loop. All gauge dynamics on the other hand are governed by a set of geometrical evolution equations, the Makeenko-Migdal equations \cite{MMeqs}:
\begin{equation}
	\partial^\nu \frac{\delta}{\delta \sigma_{\mu\nu}(x)} \mathcal{W}_1(C) =
		g^2 N_c \oint_C \Dif z^\mu \delta^{(4)}\left(x-z\right) \mathcal{W}_2(C_{xz} \, C_{zx}).
\end{equation}
Of special importance are the two geometrical operations introduced in the Makeenko-Migdal equations, namely the path derivative $\partial_\mu$ and the area derivative $\frac{\delta}{\delta\sigma_{\mu\nu}(x)}$ \cite{MMeqs}:
\begin{align}
	\partial_\mu \Phi(C) &=
		\lim_{\left|\delta x_\mu\right|\rightarrow 0}
		\frac{\Phi(\delta x_\mu^{-1} C \, \delta x_\mu) - \Phi(C)}{\left|\delta x_\mu\right|},\\
	\frac{\delta}{\delta\sigma_{\mu\nu}(x)} \Phi(C) &=
		\lim_{\left|\delta \sigma_{\mu\nu}(x)\right|\rightarrow 0}
		\frac{\Phi(C \, \delta C) - \Phi(C)}{\left|\delta \sigma_{\mu\nu}(x)\right|}.
\end{align}
There are a few limitations to the Makeenko-Migdal equations. They are not closed since the evolution of $\mathcal{W}_1$ depends on $\mathcal{W}_2$.
But more severe, the evolution equations are derived by applying the Schwinger-Dyson methodology on the Mandelstam formula
\begin{equation}
	\frac{\delta}{\delta \sigma_{\mu\nu} (x)} \Phi(C) =
		i g \,\text{tr}\, \! \left\{ F^{\mu\nu} \Phi(C_x) \right\}
\end{equation}
and using the Stokes' theorem. It is not certain that these, as well as the area derivative, are well-defined for any type of path. In particular contours containing cusps can lead to problematic behaviour, as it is not straightforward to define \emph{continuous} area differentiation inside a cusp, nor it is to continuously deform a contour in a general topology \cite{Tomspaper1,Tomspaper2,ourpaper6,ourpaper5,ourpaper4,ourpaper3,ourpaper2,ourpaper1}.

\section{Evolution of rectangular Wilson loops}
\begin{figure}[h!]
	\label{fig:rect}
	\centering
	\begin{tikzpicture}[scale=0.9]
		\draw[double,double distance=1.5pt] (0,0) -- (1,2) -- (3,2) -- (2,0) -- cycle;
		\path[very thick, postaction={decorate}, decoration={markings,mark=between positions 0.15 and 1 step .25 with {\arrow{>}}}] (0,0) -- (1,2) -- (3,2) -- (2,0) -- cycle;
		\node at (-0.25,-0.25) {$x_1$};
		\node at (0.75,2.25) {$x_2$};
		\node at (3.25,2.25) {$x_3$};
		\node at (2.25,-0.25) {$x_4$};
		\node at (0.15,1) {$v_1$};
		\node at (1.9,2.35) {$v_2$};
		\node at (2.9,1) {$v_3$};
		\node at (1.1,-0.35) {$v_4$};
		\filldraw (0,0) circle(2pt);
		\filldraw (1,2) circle(2pt);
		\filldraw (3,2) circle(2pt);
		\filldraw (2,0) circle(2pt);
	\end{tikzpicture}
	\caption{Parametrisation of a rectangular Wilson loop in coordinate space.}
\end{figure}
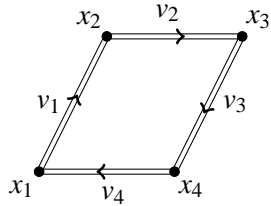
In order to avoid using non-rigorously defined methods to describe the evolution of Wilson loops, we turn our attention to different classes of contours, based on their geometrical structure, and treat the evolution on a case-by-case basis. In this paper, we will investigate a rectangular contour with light-like segments on the null-plane, as depicted in fig. (1). We aim to relate its evolution to a geometric evolution, inspired on previous work by \cite{SYM}.
The segment lengths $v_i = x_i - x_{i+1}$ are light-like ($v_i^2 = 0$) and on the null-plane ($\bvec{v}_\perp=0$). To investigate its singularity structure, we evaluate the loop \eqref{eq:wloop} at one loop level in coordinate space \cite{Drummond:2007aua}:
\begin{equation}
	\mathcal{W}_{\text{L.O.}} =
		1 -\frac{\alpha_s C_F}{\pi} \left(2\pi \mu^2 \right)^\epsilon \Gamma (1-\epsilon) \left[
			\frac{1}{\epsilon^2} \left(-\frac{s}{2} \right)^\epsilon + \frac{1}{\epsilon^2} \left(-\frac{t}{2} \right)^\epsilon
			-\frac{1}{2}\ln^2\frac{s}{t}
		\right]
\end{equation}
where $s$ and $t$ are the Mandelstam energy/rapidity variables. Then the evolution at leading order is:
\begin{equation}\label{eq:evolution}
	\frac{\Dif}{\Dif \ln \mu}\frac{\Dif}{\Dif \ln s} \mathcal{W}_{\text{L.O}} =
		-2\frac{\alpha_s C_F}{\pi} = -2 \Gamma_{\text{cusp}}
\end{equation}
where we recognise the cusp anomalous dimension in the light-cone limit from \eqref{eq:cad}. Next we introduce the area variable $ \Sigma \equiv v^-\cdot v^+ = \frac{1}{2} s$, $\frac{\delta}{\delta\ln\Sigma} = \sigma_{\mu\nu}\frac{\delta}{\delta\sigma_{\mu\nu}}$.
Replacing $s$ by $\Sigma$ in equation (\ref{eq:evolution}) gives $-4\Gamma_{\text{cusp}}$.
Motivated by this, we conjecture a general evolution equation for light-like polygon Wilson loops on the null-plane:
\begin{equation}\label{eq:result}
\frac{\Dif}{\Dif\ln\mu} \left[\sigma_{\mu\nu}\frac{\delta}{\delta \sigma_{\mu\nu}} \ln \mathcal{W} \right] =
						-\sum_i \Gamma_{\text{cusp}}.
\end{equation}
Note that this equation is in perfect agreement with the non-Abelian exponentiation theorem:
\begin{equation}
	\mathcal{W} =
		\text{exp}\left[ \sum_{k=1} \alpha_s^k C_k \left(\mathcal{W}\right) F_k \left(\mathcal{W}\right) \right]
\end{equation}
where $C_k\sim C_F N_c^{k-1}$ and the summation goes over all `webs' $F_k$, see \cite{webs}.

\section{Relation to TMDs}
Besides for light-like rectangular Wilson loops, equation (\ref{eq:result}) is expected to be valid for light-like TMDs, as they possess the same singularity structure. The area variable then gets replaced by the rapidity variable $\theta$. This gives
\begin{equation}
	\frac{\Dif}{\Dif \ln \mu}\frac{\Dif}{\Dif \ln \theta} \ln f(x,\bvec{k}_{\perp}) = 2\Gamma_{\text{cusp}}.
\end{equation}
The minus disappeared because $\theta\sim \Sigma^{-1}$, and there is a factor of 2 since we have two (hidden) cusps. Note that this result is very similar to the Collins-Soper evolution equations for off-LC TMDs. 
We can use the derived formula to get evolution equations for other similar objects, like the $\Pi$-shape Wilson (semi-)loop which has the evolution
\begin{equation}
	\frac{\Dif}{\Dif \ln \mu}\frac{\Dif}{\Dif \ln \sigma} \ln \mathcal{W}_\Pi = -2\Gamma_{\text{cusp}},
\end{equation}
which is consistent with its two cusps (see \cite{ourpaper2} for a more detailed analysis). Another example is the evolution of TMDs at large-$x_b$, see \cite{ourpaper6}.

\acknowledgments
I'd like to thank I.O.~Cherednikov, T.~Mertens and P.~Taels for our conjoint research and for their inspiring discussions. I'd also like to thank the organisers of Hadron 2013 for their hospitality and for the wonderful atmosphere during the conference.

\end{document}